# Gauge Fixing and the Gibbs Phenomenon

Jeffrey E. Mandula

U. S. Department of Energy, Division of High Energy Physics
Washington, DC 20585, United States

We address the question of why global gauge fixing, specifically to the lattice Landau gauge, becomes an extremely lengthy process for large lattices. We construct an artificial "gauge-fixing" problem which has the essential features encountered in actuality. In the limit in which the size of the system to be gauge fixed becomes infinite, the problem becomes equivalent to finding a series expansion in functions which are related to the Jacobi polynomials. The series converges slowly, as expected. It also converges non-uniformly, which is an observed characteristic of gauge fixing. In the limiting example, the non-uniformity arises through the Gibbs phenomenon.

## 1. INTRODUCTION

Although the original proposal and formulation of Lattice Gauge Theory eschewed the choice of a gauge, and such a choice is emphatically not needed for spectrum calculations, there are many contexts in which fixing the gauge on the lattice is necessary, or at least useful. Prominent among these are the use of lattice analysis to study Green's functions and the mathematical structure of field theories, as well as to evaluate the non-perturbative renormalization of operators that enter into weak matrix elements and $K^0 \bar{K}^0$ and $B^0 \bar{B}^0$ mixing coefficients.

It is well known that gauge fixing is afflicted with a host of problems. On a fundamental level the existence of Gribov copies — both the analogues of continuum copies and new lattice artifacts — remains a serious issue. We continue to lack a good understanding of how to treat them in principle, and we also have no general understanding of their practical significance, although in some examples the manner with which are dealt with or not affects the results of calculations. Presumably performing the path integral for a gauge theory within a "Fundamental Modular Domain" is an appropriate method, but we have no operational procedure for finding such a domain.[1]

Gauge fixing to the Landau gauge is also a notoriously slow process, especially for large lattices. While there are a number of techniques that work adequately well on small or moderate sized lattices, for the size lattices currently in common use, all algorithms are unfortunately very inefficient.[2] The goal of the work described here was to get some insight into this algorithmic inefficiency.

## 2. THE LATTICE LANDAU GAUGE

On the lattice, a general gauge condition $f(U) = 0$ is implemented following Fade'ev and Popov

$$\int DU\, e^{-S}\, O(U) \Big|_{f(U)=0}$$
$$= \int DU \int DG\, e^{-S}\, O(U)\, \Delta_{FP}(U)\, \delta(f(U^{(G)}))$$
$$= \int DU\, e^{-S}\, O(U^{(G)})$$

where $U^{(G)}$ is the gauge transform of $U$, satisfying $f(U^{(G)}) = 0$.

For the lattice Landau gauge, one formulates the gauge condition as a maximization so as to avoid some lattice artifacts

$$\underset{G}{Max} \sum_{x\mu} Re\, Tr\, U_\mu^{(G)}(x)$$

The most naive implementation is to cycle through the lattice, imposing the maximization requirement one site at a time, and repeating the process until the gauge transform has relaxed sufficiently into a global minimum.[3] Two classes of improvements on this method in practical use are 1) Overrelaxation[4], the replacement $G_{max}(x) \rightarrow G_{max}^w(x)$, $(1 \leq w \leq 2)$, either exactly or stochastically, and 2) Fourier Preconditioning[5].

The conventional wisdom about the inefficiency of gauge fixing is that the lattice configuration develops "hot spots", that is exceptional poorly fixed points which move from site to site under the effect of repeated gauge fixing sweeps, but which are strongly persistent. Another part of the conventional wisdom, not actually compatible with the idea of "hot spots", is that it is the longest wavelength modes that relax most slowly.

If one looks in detail at the distribution of values of some measure of the deviation from Landau gauge at each site, $e.g.$, $Tr(\Delta_\mu A_\mu)^2$ one sees not just a few very poorly fixed sites, but a broad range of deviations with no notable gaps. If one looks in Fourier space, one finds that there is a broad range of relaxation times, with many modes decaying slowly, including, but not limited to, the longest wavelengths.

## 3. CONJUGATE GRADIENT GAUGE FIXING

Since Landau gauge fixing is a maximization problem, one can try to attack it with the standard tools for such. Conjugate gradient is often the best such. In lattice applications it is most familiar as a technique for inverting the Dirac operator, in fact it is a general method for finding the minimum or maximum of a multivariate function.

If the link variables $U_\mu(x)$ are close to satisfying the Landau gauge condition, we may expand $G(x) = \exp(i\, g(x))$ to second order, and express the Landau gauge condition as the maximization of a quadratic form

$$\underset{g(x)}{Max}\ Re\ Tr\, i\, [g(x)\, U_\mu(x) - U_\mu(x - \hat{\mu})\, g(x)]$$

$$- \frac{1}{2} Re\ Tr\, [g^2(x)\, U_\mu(x) - U_\mu(x - \hat{\mu})\, g^2(x)]$$

$$+ Re\ Tr\, [g(x)\, U_\mu(x)\, g(x + \hat{\mu})]$$

Setting the variation of the quadratic form with respect to the generator of gauge transformations, $g(x)$, to zero, gives a matrix inversion problem with a nearest neighbor structure. The eigenvalues of this matrix control the convergence of relaxation methods, and a two dimensional example illustrates the typical character of the eigenvalues.

The eigenvalues of the tridiagonal $N \times N$ matrix

$$M_{i,i+1} = M_{i+1,i} = M_{1,1} = M_{N,N} = 1$$

$$M_{i,i} = 2 \qquad i = 2, ..., N-2$$

can be found analytically, and small eigenvalues are

$$\lambda_m \cong -\frac{m^2 \pi^2}{2N^2} \qquad\qquad m \ll N$$

The important points to note are that there is always a zero eigenvalue, and that as $N$ becomes large, many small eigenvalues develop. As we shall see, it is not the single vanishing eigenvalue which is troublesome, but the multiplicity of small ones.

If $M$ is the quadratic form to be extremized, conjugate gradient is a recipe for finding a sequence of mutually conjugate vectors, that is, vectors satisfying

$$(h_i, M h_j) = 0 \qquad (i \neq j)$$

Each vector the sequence constructed from its predecessors

$$h_{i+1} = M h_i + \sum_{j=0}^{j=i} c_j h_j$$

The initial vector is taken to be in direction of steepest descent from the starting point.

Let us study the convergence of conjugate gradient minimization in a simple model with all its characteristic features: The minimization of an $N$-dimensional quadratic form with eigenvalues evenly distributed between 0 and 1, choosing as a starting point a vector with equal projection on each eigenvector. The handiest basis in which to analyze the problem is that in which the matrix is diagonal:

$$Min \sum_{i,j=1}^{N} v_j M_{ij} v_j \qquad M_{ij} = \lambda_i \delta_{ij} = \frac{i}{N} \delta_{ij}$$

In this basis the starting position is

$$h_i^{[0]} = 1 \qquad i = 1, ..., N$$

and the sequence of conjugate gradient directions follows mechanically.

## 4. THE N → ∞ LIMIT

If we pass to the $N \to \infty$ limit, rescale index to a continuous variable on the interval, and replace sums by integrals, then the mutually conjugate vectors are replaced by polynomials $P_n(x)$, which mutually orthogonal with respect to the weight function $w(x) = x$ on the interval $x \in [0,1]$. The initial point, which we may take as the conventional origin, is the constant function. With suitable normalizations, the convergence of the minimization procedure is equivalent to the convergence of the sequence of approximations

$$f_n(x) \equiv \sum_{m=1}^{n} c_m P_m(x) \to f(x) = 1$$

The polynomials $P_n(x)$ are related to Jacobi polynomials, and one can work out all their properties and expansions using standard analysis methods.

Specifically, the $P_n(x)$ are the regular solutions of

$$\left[\frac{d}{dx} x^2(1-x)\frac{d}{dx} - 2\right] P_n(x) = -n(n+2)x P_n(x)$$

A convenient normalization for these polynomials is

$$P_n(x) = \sum_{m=1}^{n} \frac{(-1)^{m-1}(n-1)!}{(m-1)!(n-m)!} \frac{3!(n+m+1)!}{(n+2)!(m+2)!} x^m$$

The approximations $f_n(x)$ are quite simply expressed

$$f_n(x) = \frac{2}{3} \sum_{m=1}^{n} (n+1) P_m(x)$$

$$= 1 - 2 \sum_{m=0}^{n} \frac{(n!)^2 (-x)^m (1-x)^{n-m}}{m!(m+2)!((n-m)!)^2}$$

The convergence of the partial sums in the norm is only a power of the number of terms,

$$\int_0^1 x (f_n(x) - 1)^2 = \frac{2}{[(n+1)(n+2)]^2}$$

so that the relaxation time of the sequence is infinite.

The pointwise convergence is both slow and non-uniform. At fixed $x$ in the interior of $(0,1)$ one has

$$f_n(x) - 1 \sim \frac{\cos(n\theta + c)}{n^{-5/2}} \quad (\cos\theta = 1 - 2x)$$

However, at the upper end of the interval,

$$f_n(1) = \frac{2(-1)^n}{(n+1)(n+2)}$$

which is a slower rate of convergence. At the lower end of the interval, pointwise convergence actually fails. For small values of $x \propto 1/n^2$, deviation from unity remains finite for all $n$:

$$\lim_{n \to \infty} f_n\left(\frac{a}{n^2}\right) = 1 - \frac{2}{a} J_2(2\sqrt{a})$$

where, $J_2$ denotes the Bessel function. The maximum value of this expression is about 1.059, which is attained at $a \approx 1.8$

This is an example of the Gibbs phenomenon. The series overshoots and oscillates about true value. Even though this behavior is restricted to an ever smaller range, $x \propto 1/n^2$, it saturates the convergence in the norm.

The Gibbs phenomenon and all the other convergence difficulties discussed come from the multiplicity of small eigenvalues. If we replaced the polynomials $P_n(x)$ with polynomials $K_n(x)$, taken to be orthogonal with respect to a weight function that does not take arbitrarily small values, say

$$w(x) = x \to w(x) = (x+1)$$

the expansion would converge exponentially with $n$.